\documentclass[aps,showpacs,amsmath,amssymb]{revtex4}
\usepackage{graphicx}
\usepackage{dcolumn}
\usepackage{bm}
\usepackage{epsfig}
\usepackage{multirow}
\usepackage{booktabs}

\begin{document}
\title{Quasi-normal modes of the Ayón–Beato–García black hole surrounded by quintessence}

\author{Diego Ariel Sotelo Carrillo$^1$}
\email{diegoariel.sotelocarrillo@fisica.uaz.edu.mx}
\author{Omar Pedraza$^2$}
\email{omarp@uaeh.edu.mx}
\author{L. A. López$^2$}
\email{lalopez@uaeh.edu.mx}
\author{R. Arceo$^3$}
\email{roberto.arceo@unach.mx}

\affiliation{$^{1}$ 
Campus Universitario II Avenida Solidaridad, Hidráulica,  Universidad Aut\'onoma de Zacatecas, C. P. 98068, Zacatecas, Zacatecas, M\'exico.}

\affiliation{$^2$ \'Area Acad\'emica de Matem\'aticas y F\'isica, UAEH, 
	Carretera Pachuca-Tulancingo Km. 4.5, C. P. 42184, Mineral de la Reforma, Hidalgo, M\'exico.}

\affiliation{$^{3}$ Facultad de Ciencias en F\'isica y Matem\'aticas, Universidad Aut\'onoma de Chiapas, C. P. 29050, Tuxtla Guti\'errez, Chiapas, M\'exico.}

\begin{abstract}

We compute the quasi-normal modes of scalar, electromagnetic, and gravitational perturbations of an Ay\'on–Beato–Garc\'ia black hole surrounded by quintessence using the Asymptotic Iteration Method (30 iterations). The results are compared with the WKB approximation to evaluate the robustness and accuracy of AIM. The results show that the real part of the frequencies increases with the charge parameter, while the absolute value of the imaginary part decreases as the charge parameter increases. In contrast, both the real part and the absolute value of the imaginary part decrease as the quintessence parameter $c$ increases. These trends are observed for both values of the quintessence state parameter, $\omega_q = -2/3$ and $\omega_q = -4/9$; however, the effects become more pronounced as $\omega_q$ approaches $-1$ corresponding to cosmological constant. The quasi-normal frequencies tend to have finite values, which could facilitate their observational detection. Moreover, the presence of quintessence leads to a slower damping of the perturbations.
\\
\\
{\it Keywords:} Regular black holes, quasi--normal modes, quintessence, AIM.
\pacs{04.20.-q, 04.70.-s, 04.70.Bw, 04.20.Dw}
\end{abstract}

\maketitle
\section{Introduction}

The study of quasi-normal modes (QNMs) associated with black holes plays a central role in characterizing their stability and understanding the emission of gravitational waves. These modes depend on the fundamental parameters that define the black hole and are described by a Schrödinger-type wave equation with a specific effective potential. The boundary conditions, which require purely ingoing waves at the event horizon and outgoing waves at spatial infinity, lead to complex oscillatory solutions. The real part of the frequency corresponds to the oscillation rate, while the imaginary part determines the damping time scale.

The study of quasi-normal modes has been carried out for a wide range of regular and no regular black hole solutions, most of which are typically treated as isolated systems among those that can be mentioned, for example \cite{Lopez:2022uie} \cite{Toshmatov:2018tyo} \cite{Fernando:2012yw}. However, this assumption is not entirely realistic, since black holes are astrophysical objects distributed throughout the universe and can interact with different types of matter and fields. Considering that dark energy constitutes approximately $68 \%$ of the universe and is responsible for its accelerated expansion. It is reasonable to assume that black holes are surrounded by dark energy. In this sense, QNM studies have also been carried out on black holes surrounded by dark energy or dark matter, for example \cite{Al-Badawi:2025coy} \cite{Das:2023ess} \cite{Al-Badawi:2023lke}.

As is well known, several alternative models have been proposed as candidates for dark energy, most of which rely on the dynamics of a scalar field. Among these, the quintessence model has been extensively employed to describe black holes surrounded by such a field.

Kiselev \cite{Kiselev:2002dx} presented a static spherically symmetric solution of the Einstein equations that represent a black hole surrounding for quintessential. It is possible to mention that although the stress–energy tensor leading to Kiselev model is  anisotropic and can be decomposed into a perfect-fluid component plus an electromagnetic or scalar-field contribution implying that it does not correspond to a perfect fluid or canonical quintessence in the cosmological sense \cite{Boonserm:2019phw}  \cite{Visser:2019brz}. Nevertheless, the Kislev model remains a useful, that captures  features of dark-energy-like fields.

The Kiselev model has enabled various studies related to black holes surrounded by quintessence, such as geodesics \cite{Ghaderi:2017yfr}, quasi-normal modes \cite{Rayimbaev:2022mrk}, scattering cross-sections, and absorption \cite{Lopez:2021ujg}. Generally, it is possible to observe that an additional horizon (quintessence horizon) can be obtained when the quintessence is incorporated. Also, the quintessence state parameter and the normalization factor modify the behaviour of QNMs,  scattering cross-sections, and absorption cross-sections.

Among the different models of black holes, regular black holes have attracted considerable attention due to their ability to avoid singularities while remaining consistent with general relativity. In particular, the Ay\'on–Beato–Garc\'ia (ABG) solution \cite{ayon1998regular} stands out as an appealing framework, since it arises from nonlinear electrodynamics coupled to general relativity and describes a nonsingular space-time. Over the years, several properties of the ABG black hole have been extensively studied, including the motion of test particles \cite{Garcia:2013zud}, the spectrum of quasi-normal modes \cite{Toshmatov:2015wga}, absorption phenomena, superradiant amplification effects \cite{dePaula:2024xnd}, and its thermodynamic stability \cite{Lopez:2016oaw}. 

In this work, we investigate the quasi-normal modes of scalar, electromagnetic, and gravitational perturbations of the Ayón–Beato–García black hole surrounded by quintessence. To this end, we employ the Asymptotic Iteration Method (AIM), an efficient and accurate numerical technique for solving Schrödinger-type differential equations, and compare the resulting frequencies with those obtained via the WKB approximation to assess the reliability of the AIM results. This framework allows us to compute the complex QNM spectrum and examine how both the quintessence and the  black hole parameters modify the dynamics of perturbations.

This paper is organized as follows. Section \ref{sec.hori} briefly describes the Ayón-Beato-García black holes surrounded by quintessence and analyzes their event horizons. Specifically, we examine the parameter regions where the black holes exhibit three, two, or one horizon. In Section \ref{sec.ep}, we describe the scalar, electromagnetic, and gravitational perturbations of the black hole. In Section \ref{sec.qnm}, we present the results for the quasi-normal modes using the AIM and WKB. Finally, we summarize our conclusions in Section \ref{conclu}.

\section{Ay\'on--Beato--Garc\'ia black hole surrounded by quintessence}\label{sec.hori}

The Ay\'on--Beato--Garc\'ia regular solution \cite{ayon1998regular} that describe a black hole in general
relativity coupled to non-linear electrodynamics and  that satisfies the weak energy condition has the line element; 

\begin{equation}
ds^2=-f(r)dt^2+\frac{dr^2}{f(r)}+r^2\left(d\theta^2+\sin^2\theta d\phi^2\right)\,,
\end{equation} 
where
\begin{equation}\label{fm}
f(r)=1-\frac{2Mr^2}{\left(r^2+g^2\right)^{3/2}}+\frac{g^2r^2}{\left(r^2+g^2\right)^2}\,,
\end{equation}
where $M$ is the mass of the black hole (BH), and $g$ denotes the charge parameter. Asymptotically, (\ref{fm}) behaves as Reissner–Nordstrom a black hole. 

On the other hand, it is well known that Kiselev \cite{Kiselev:2002dx} proposed a model describing black holes surrounded by a quintessence field.

Following this approach, the quintessence field is characterized by the following relations:

\begin{equation}\label{key}
T^{\phi}_{\phi},\quad T_{\theta}^{\theta}=-\frac{1}{2}\left(3\omega_q+1\right)T_r^r=\frac{1}{2}\left(3\omega_q+1\right)T_t^t\,,
\end{equation} 
where $\omega_q$ is the quintessence state parameter which has the range $-1<\omega_q<-1/3$. 

The energy density $\rho$ is given by
\begin{equation}\label{ec.cr}
\rho=T_{tt}=-\frac{c}{2}\frac{3\omega_q}{r^{3\left(\omega_q+1\right)}}\,.
\end{equation}
The quintessence pressure ($p=\rho\omega_q$) must be negative in order to drive the accelerated expansion of the universe. Since the energy density must be positive, it follows from equation \eqref{ec.cr} that the normalization factor $c$ must also be positive for any negative value of $\omega_q$. 

Following the ideas mentioned above, the metric function of a BH surrounded by quintessence can be obtained by adding the quintessence term $-c/r^{3\omega_q+1}$ to the BH metric (see, for example, \cite{Saleh:2018hba},\cite{Pedraza:2021hzw},\cite{AlBadawi:2025coy},\cite{Hamil:2024njs}). Thus, the metric function for a ABG black hole surrounded by quintessence can be written as
\begin{equation}\label{fm}
	f_{\omega_q}(r)=1-\frac{2Mr^2}{\left(r^2+g^2\right)^{3/2}}+\frac{g^2r^2}{\left(r^2+g^2\right)^2}-\frac{c}{r^{3\omega_q+1}}\,.
\end{equation}

When the quintessence term is taken into account, an additional horizon appears, commonly referred to as the cosmological (quintessence) horizon. Then the number of horizons of Ay\'on--Beato--Garc\'ia black hole surrounded by quintessence depends entirely on the  values of the different parameters. 
To carry out the analysis, it is crucial to determine the event horizons. This involves examining the positive roots of the equation $f_{\omega_q}\left(r_h\right)=0$, where $r_h$ denotes the radius of the horizon. Imposing this condition leads to the following expression:

\begin{equation}\label{ec.rh}
\left(r^{3\omega_q+1}_h-c\right)\left(g^2+r^2_h\right)^2-2\left(g^2+r^2_h\right)^{1/2} r^{3\omega_q+3}_h+g^2 r^{3\omega_q+3}_h=0\,.
\end{equation}
Here, the parameters $g$, $c$ and the radial coordinate $r$ have been expressed in units of the black hole mass $M$, such that $g \to g/M$, $c \to c/M^{3\omega_q + 1}$ and $r \to r/M$. Depending on the values of $(g^2, c)$, the space-time may represent a black hole, an extremal black hole, or a black hole with a single apparent horizon.

From \eqref{ec.rh}, we can parametrize $c$ as a function of $g^2$ and $r_h$ as
\begin{equation}\label{ec.exc}
c\left(g^2,r_h\right)=r^{3\omega_q+1}_h\frac{\left(g^2+r^2_h\right)^2+g^2r^2_h-2r^2_h\left(g^2+r^2_h\right)^{1/2}}{\left(g^2+r^2_h\right)^2}\,.
\end{equation}

Using the method described in \cite{Lopez:2021ujg}, we obtain the critical values of $g^2_c$ and  $c_c$, as shown in Table \ref{Ta1}.

\begin{table}[!hbt]
	\begin{center}
		\begin{tabular}{|c|c|c|}
			\hline
			$\omega_q$ & $-4/9$ &  $-2/3$ \\
			\hline
			$g^2_c$ &1.4042 & 0.6235 \\
			\hline
			$c_c$ & 0.3990&  0.1559  \\
			\hline
		\end{tabular}
		\caption{Critical values of the $g^2$ and $c$ for different $\omega_q$ (see also Fig.  \ref{frvsg}(b)).}
		\label{Ta1}
	\end{center}
\end{table}

Following the approach outlined in Ref. \cite{Rizwan:2018lht}, the condition for the existence of an extremal black hole can be obtained for any value of $\omega_q$ from the simultaneous satisfaction of the following two equations
\begin{eqnarray}\label{key}
\frac{df_{\omega_q}(r)}{dr}&=&
\frac{
	4r_h^{3\omega_q+3}\left(r_h^2+g^2\right)^{3/2}-6r_h^{3\omega_q+5}\left(r_h^2+g^2\right)^{1/2}-c\left(3\omega_q+1\right)\left(r_h^2+g^2\right)^{3}-2g^2r_h^{3\omega_q+3}\left(g^2-r_h^2\right)
}{\left(r_h^2+g^2\right)^3r_h^{3\omega_q+2}}=0\,,\\
f_{\omega_q}(r) &=& 0\,.
\end{eqnarray}
By combining both conditions, we obtain the following expression
\begin{equation}\label{ec.ehe}
\left[4g^2r_h^2-2r_h^4-\left(3\omega_q+1\right)\left[\left(r_h^2+g^2\right)^2+g^2r_h^2-2r_h^2\left(r_h^2+g^2\right)^{1/2}\right]\left(r_h^2+g^2\right)^{1/2}
\right]\left(r_h^2+g^2\right)^{1/2}
-2g^2r_h^2\left(g^2-r_h^2\right)=0\,.
\end{equation}

\begin{figure}[h!]
	\centering
	\includegraphics[scale=0.85]{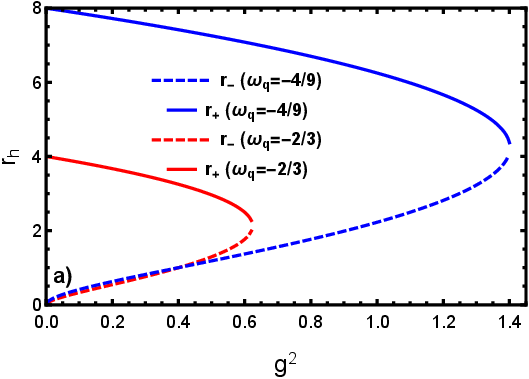}
	\includegraphics[scale=0.87]{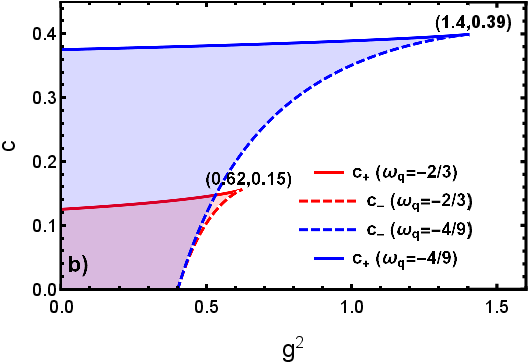}
	\caption{(a) The plot shows the behavior of $r_h$ as function of $g^2$ for $\omega_q = -2/3$ and $\omega_q = -4/9$. (b) The plot illustrates how $c$ varies with $g^2$ for the cases $\omega_q = -2/3$ and $\omega_q = -4/9$.}
	\label{frvsg}
\end{figure}

The next step is to solve numerically \eqref{ec.ehe} for two particular cases: $\omega_q = -2/3$ and $\omega_q = -4/9$. From the numerical analysis of \eqref{ec.ehe}, it can be seen that there are only two real positive roots, $r_+$ and $r_-$, for any value of $\omega_q$. The Fig. \ref{frvsg} a) shows the behavior of the $r_+$ and $r_-$ for $\omega_q = -2/3$ and $\omega_q = -4/9$. 

Using the values obtained from the numerical analysis of equation \eqref{ec.ehe} in \eqref{ec.exc}, it is possible to observe the behavior of the quintessence parameter $c$ as a function of $g^2$, as shown in Fig. \ref{frvsg} b). This plot helps determine the number of horizons of ABG black hole surrounded by quintessence. Specifically, for values of $(g^2, c)$ lying between the curves $c_+$ and $c_-$ (the red area corresponds to $\omega_q = -2/3$, and the blue area corresponds to $\omega_q = -4/9$) in Fig. \ref{frvsg} b), there are three real positive roots of Eq. \eqref{ec.rh}: $r_{\text{in}}$, $r_{\text{out}}$, and $r_{\omega_q}$, which satisfy the relation $r_{\text{in}} \leq r_{\text{out}} \leq r_{\omega_q}$.

For values of $(g^2, c)$ lying on the curves $c_+$ or $c_-$, the solution corresponds to an extremal black hole. In contrast, in the regions outside these curves $c_+$ and $c_-$ (i.e., for $c \leq c_c$ and $g^2 \leq g_c^2$), there exists only the quintessence horizon.

Again, from Fig. \ref{frvsg}(b), by comparing the regions between the blue and red curves, it can be seen that  the area enclosed by the blue region ($\omega_q=-4/9$) is larger than that of the red region ($\omega_q=-2/3$). While the values of the BH parameters $(g,c)$ lying on the curves $c_+$ and $c_-$ imply that the BH has two horizons.

\section{Massless scalar, electromagnetic and gravitational perturbations}\label{sec.ep}

When a BH is perturbed, its response is characterized by damped oscillations with complex frequencies known as QNMs. The study of QNMs reduces to solving a Schrödinger-like wave equation with an effective potential determined by the  metric function. For static and spherically symmetric space-times, one can take as a starting point the Schr\"odinger-type equation for a scalar function $R(r)$ given by:

\begin{equation}\label{ec.ts}
\frac{d^2R(r)}{dr_*^2}+\left[\omega^2-V_{i}(r)\right]R(r)=0\,,
\end{equation}
 
 Here the $r_*$ is the tortoise coordinate, defined by
\begin{equation}
dr_*=\frac{dr}{f_{\omega_q}(r)}\,.
\end{equation} and the effective potential $V_{i}(r)$ is given by,

\begin{equation}\label{ec.ps}
V_{i}(r)=f_{\omega_q}(r)\left[\frac{l(l+1)}{r^2}+\left(1-i^2\right)\frac{2m(r)}{r^3}
+\left(1-i\right)\left(\frac{f'_{\omega_q}(r)}{r}-\frac{2m(r)}{r^3}
\right)
\right],
\end{equation}

and $m(r)$ as:

\begin{equation}
m(r)=\frac{Mr^3}{\left(r^2+g^2\right)^{3/2}}
	-\frac{g^2r^3}{2\left(r^2+g^2\right)^2}+\frac{c}{2r^{3\omega_q}}\,,
\end{equation}

where $l$ is the spherical harmonic index restricted by $l\geq i$, and $i=0,1,2$  denotes the spin of the perturbation, corresponding to scalar, electromagnetic, and gravitational fields, respectively. 

\begin{figure}[h!]
	\centering
	\includegraphics[scale=0.87]{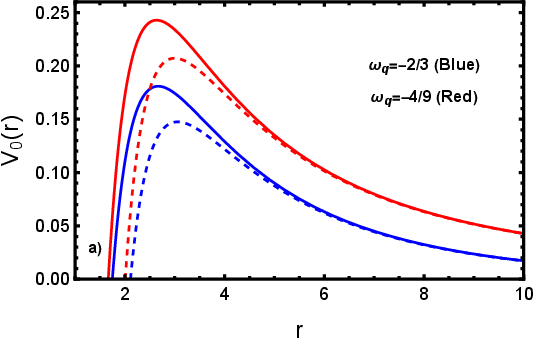}
	\includegraphics[scale=0.87]{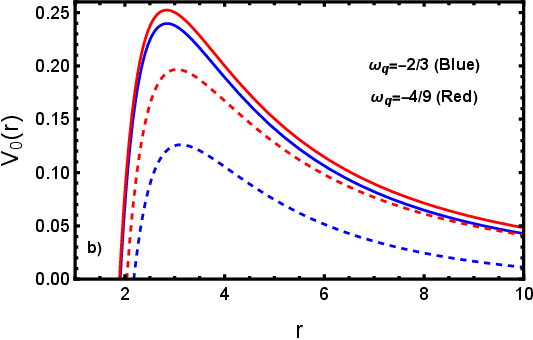}
	\caption{a) The behavior of the scalar effective potential $V_0(r)$ is shown for different values of $g^2$, with $c = 0.05$ and $l = 2$. The solid line corresponds to $g^2 = 0.3$, while the dashed line represents $g^2 = 0.1$.
		b) The behavior of the scalar effective potential $V_0(r)$ is shown for different values of $c$, with $g^2 = 0.1$ and $l = 2$. The solid line corresponds to $c = 0.01$, while the dashed line represents $c = 0.06$.}
	\label{vvsgc}
\end{figure}

\begin{figure}[h!]
	\centering
	\includegraphics[scale=0.87]{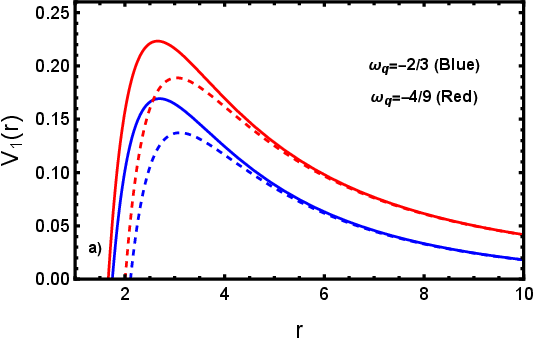}
	\includegraphics[scale=0.87]{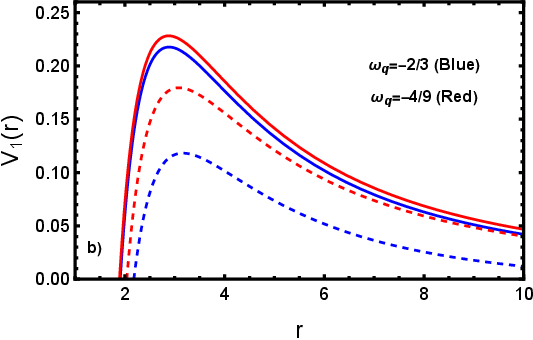}
	\caption{a) Electromagnetic effective potential $V_1(r)$ for different values of $g^2$, with $c = 0.05$ and $l = 2$. The solid and dashed lines correspond to $g^2 = 0.3$ and $g^2 = 0.1$, respectively.
		b) Electromagnetic effective potential $V_1(r)$ for different values of $c$, with $g^2 = 0.1$ and $l = 2$. The solid and dashed lines correspond to $c = 0.01$ and $c = 0.06$, respectively.}
	\label{vvsgce}
\end{figure}
\begin{figure}[h!]
	\centering
	\includegraphics[scale=0.87]{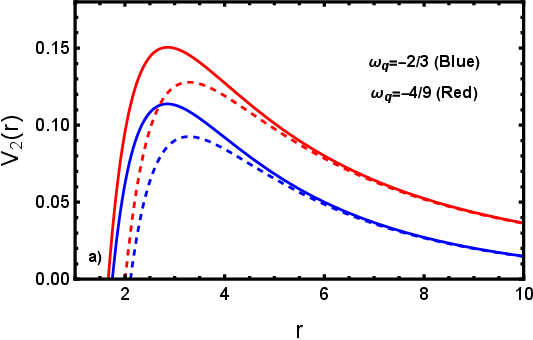}
	\includegraphics[scale=0.87]{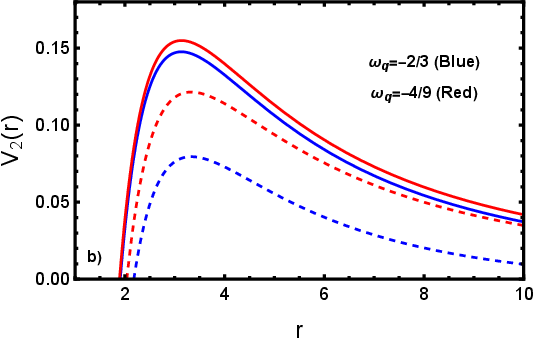}
	\caption{a) Gravitational effective potential $V_1(r)$ for different values of $g^2$, with $c = 0.05$ and $l = 2$. The solid and dashed lines correspond to $g^2 = 0.3$ and $g^2 = 0.1$, respectively.
		b) Gravitational effective potential $V_1(r)$ for different values of $c$, with $g^2 = 0.1$ and $l = 2$. The solid and dashed lines correspond to $c = 0.01$ and $c = 0.06$, respectively.}
	\label{vvsgcb}
\end{figure}

The Fig.~\ref{vvsgc} illustrates the scalar effective potential $V_{0}(r)$ (\ref{ec.ps}) associated with scalar perturbations for  $\omega = -2/3$ and $\omega = -4/9$. It is observed that, for a fixed value of the parameter $c$, the height of the potential barrier increases as the charge parameter $g^{2}$ becomes larger (see Fig.~\ref{vvsgc} a)). Furthermore, when the parameter $c$ is reduced, the  amplitude of the potential decreases and the location of its maximum is displaced toward  the right (see Fig.~\ref{vvsgc} b)).

In Fig.~\ref{vvsgce} a), we present the behavior of the electromagnetic potential $V_{1}(r)$ (\ref{ec.ps}) for $\omega_q = -4/9$ and $\omega_q = -2/3$ as the parameter $g^{2}$ varies. For both values of $\omega_q$, the height of the potential barrier decreases as $c$  increases (see Fig. \ref{vvsgce} b)), indicating that the presence of quintessence suppresses the magnitude of the effective potential. Furthermore, Figs.~\ref{vvsgce} a) and \ref{vvsgce} b) reveal that the maximum of the effective potential grows as the state parameter $\omega_q$ decreases. The same behaviors can be observed for the case of gravitational perturbations $V_{2}(r)$ (\ref{ec.ps}) (see Figs ~\ref{vvsgcb} a) and \ref{vvsgcb} b)). 

In general, the different figures indicate that the effective potential satisfies $V(r)_{-4/9} > V(r)_{-2/3}$ for all types of perturbations considered, consequently, as $\omega_q \rightarrow -1$, the difference between the corresponding potentials is expected to become even more pronounced. Moreover, we observe that $V_{0}(r)> V_{1}(r)>V_{2}(r)$ which indicates that scalar perturbations produce a strongest response in the ABG black hole surrounded by quintessence. The effects of incorporating quintessence are consistent with those reported for other black hole solutions \cite{Pedraza:2021hzw} surrounded by a quintessence field.

\section{Quasi--normal modes}\label{sec.qnm}

In this next section, we obtain numerically the spectrum of the QNMs for the background given by \eqref{fm}, considering scalar, electromagnetic, and gravitational perturbations.

To find the QNMs from Eq. (\ref{ec.ts}), we impose that the wave is purely ingoing at the event horizon and purely outgoing at spatial infinity. These boundary conditions can be expressed mathematically as:
\begin{equation}\label{ec.bca}
	R(r_*) \sim
	\begin{cases}
		e^{-i\omega r_*} \sim \left(r - r_h\right)^{-i\omega / 2\kappa_h}, & \text{as } r_* \to -\infty \; (r \to r_h), \\	
		e^{i\omega r_*} \sim e^{i\omega \int \frac{dr}{f_{\omega_q}(r)}}, & \text{as } r_* \to +\infty \; (r \to \infty),
	\end{cases}
\end{equation}
where $\kappa_h$ is the surface gravity at the event horizon, defined as
\begin{equation}\label{ec.gs}
	\kappa_h = \left. \frac{1}{2} \frac{df_{\omega_q}(r)}{dr} \right|_{r = r_h}.
\end{equation}

The QNMs are the complex frequency solutions, $\omega = \omega_r + i\omega_i$, of (\ref{ec.ts}), which arise from the interaction between the ABG black hole surrounded by quintessence and the perturbing field, and satisfy the boundary conditions given in (\ref{ec.bca}). This model type can provide valuable insights into the astrophysical properties of black holes and may play a significant role as an indirect method for detecting dark energy near black holes. Here, $\omega_{r}$ denotes the real part of the QNMs, corresponding to the oscillation frequency of the perturbation, while $\omega_{i}$ represents its imaginary part and characterizes the stability of the BH.

Several numerical and semi-analytical methods exist in the literature to compute the QNMs of black holes. For example, Mashhoon's method replaces the potential barrier in the Schr\"odinger-type equation with one that is analytically solvable, yielding accurate results for high multipole numbers. Leaver's method employs a Frobenius series expansion that leads to continued fractions for determining QNMs. The WKB approximation \cite{Iyer:1986nq}, which has been extended up to sixth order \cite{Konoplya:2003dd},\cite{Konoplya:2004ip}, offers improved accuracy but becomes unreliable when the overtone number $n$ is comparable to the angular number $l$. 

Another widely used technique is the Asymptotic Iteration Method, which in many cases provides an efficient and reliable procedure for computing QNMs. In this work we will apply the AIM, hence we provide below a brief summary of the method and its implementation for computing the QNMs.

To carry out the AIM approach, it is convenient to implement the following change of variable $r=1/\xi$. In terms of the new variable, the Eq. (\ref{ec.ts}) can now be expressed as:

\begin{equation}\label{ec.aim1}
\frac{d^2R}{d\xi^2}+\left(\frac{d}{d\xi}\ln\left[p\left(\xi\right)\right]\right)\frac{dR}{d\xi}+\left[
\frac{\omega^2}{p^2\left(\xi\right)}-\frac{l\left(l+1\right)}{p\left(\xi\right)}+\frac{1}{\xi}\left(\frac{d}{d\xi}\ln\left[p\left(\xi\right)\right]\right)-\frac{2}{\xi^2}
\right]	R=0\,,
\end{equation}
where 
\begin{equation}
p(\xi)=\xi^2-\frac{2\xi^3}{\left(1+g^2\xi^2\right)^{3/2}}+\frac{g^2\xi^4}{\left(1+g^2\xi^2\right)^2}-c\xi^{3\left(\omega_q+1\right)}\,.\label{ec.dp}
\end{equation}
To apply the boundary conditions and find the quasi--normal modes, it is necessary to express \eqref{ec.bca} in terms of the variable $\xi$. Thus, note that as $r \to r_h$, the function $R$ behaves as
\begin{equation}\label{key}
R \sim (r - r_h)^{-i\omega / 2\kappa_h} \sim (\xi - \xi_h)^{-i\omega / \kappa_h}\,,
\end{equation}
and as $r \to \infty$, we have
\begin{equation}\label{key}
R \sim e^{i\omega r_*} \sim e^{-i\omega \int \frac{d\xi}{p(\xi)}}\,.
\end{equation}
Therefore, to solve \eqref{ec.aim1} under the boundary conditions given by \eqref{ec.bca}, using the AIM, we propose the following expression
\begin{equation}\label{ec.rfr}
R(\xi)=\left(\xi-\xi_h\right)^{-i\omega/\kappa_h}e^{-i\omega\int\frac{d\xi}{p(\xi)}}\chi(\xi)\,.
\end{equation}
Introducing (\ref{ec.rfr}) into Eq. (\ref{ec.aim1}), we have the following expression
\begin{equation}\label{ec.aim}
\frac{d^2\chi(\xi)}{d\xi^2}=\lambda_0(\xi)\chi(\xi)+s_0(\xi)\chi(\xi)\,,
\end{equation}
where
\begin{eqnarray}
\lambda_0\left(\xi\right)&=&
\frac{2i\omega}{\kappa_h\left(\xi-\xi_h\right)}-\frac{d}{d\xi}\ln\left[p\left(\xi\right)\right]+\frac{2i\omega}{p\left(\xi\right)}
\label{l0} \,,\\
s_0\left(\xi\right)&=&
\frac{l\left(l+1\right)}{p\left(\xi\right)}
+\frac{2}{\xi^2}
-\frac{\frac{d}{d\xi}\ln\left[p\left(\xi\right)\right]}{\xi }
-\frac{i\omega\left(\kappa_h+i\omega\right)}{\kappa^2_h\left(\xi-\xi_h\right)^2}
+
\frac{i\omega \frac{d}{d\xi}\ln\left[p\left(\xi\right)\right]}{\kappa_h\left(\xi-\xi_h\right)}
+
\frac{2\omega^2}{p\left(\xi\right)\kappa_h\left(\xi-\xi_h\right)}
\,.\label{s0}
\end{eqnarray}

Differentiating (\ref{ec.aim}) with respect to $\xi$ iteratively, for the $(n+2)$th derivatives, then; 
   
\begin{equation}\label{ec.aim2}
\frac{d^{n+2}\chi(\xi)}{d\xi^{n+2}}=\lambda_n(\xi)\frac{d\chi(\xi)}{d\xi}+s_n(\xi)\chi(\xi)\,,
\end{equation}
where the coefficients are
\begin{eqnarray}
	\lambda_n(\xi)&=&\frac{d\lambda_{n-1}(\xi)}{d\xi}+s_{n-1}(\xi)+\lambda_0(\xi)\lambda_{n-1}(\xi)\,,\label{ec.ln0}\\
	s_n(\xi)&=&\frac{ds_{n-1}(\xi)}{d\xi}+s_0(\xi)\lambda_{n-1}(\xi)\,.\label{ec.sn0}
\end{eqnarray}
For sufficiently large $n$, the
coefficients $\lambda_n(\xi)$ and $s_n(\xi)$ satisfy the quantization relation
\begin{equation}\label{ec.qz}
	s_n\lambda_{n-1}-s_{n-1}\lambda_n=0\,.	
\end{equation}

Now, to calculate the QNMs of ABG black hole surrounded by quintessence, we follow the improved AIM procedure (Additional details can be found in Ref.\cite{Lopez:2023ixb,Cho:2011sf}). To this end, we expand expressions \eqref{ec.ln0} and \eqref{ec.sn0} in a Taylor series around the point $\xi_0$ at which the effective potential is maximized; 
\begin{eqnarray}
	\lambda_n(\xi)&=&\sum_{i=0}^{\infty}c_n^i\left(\xi-\xi_0\right)^i\,,\label{ec.ln}\\
	s_n(\xi)&=&\sum_{i=0}^{\infty}d_n^i\left(\xi-\xi_0\right)^i\,,\label{ec.sn}
\end{eqnarray} 

where $c_n^i$ and $d_n^i$ are the $i$th Taylor coefficients. Substituting (\ref{ec.ln}) and (\ref{ec.sn}) in (\ref{ec.ln0}) and (\ref{ec.sn0}), we have the following recursion relations for the coefficients.

\begin{eqnarray}
	c_n^i&=&(i+1)c_{n-1}^{i+1}+d_{n-1}^i+\sum_{j=0}^ic_0^jc_{n-1}^{i-j}\,,\label{ec.ln1}\\
	d_n^i&=&(i+1)d_{n-1}^{i+1}+\sum_{j=0}^id_0^jc_{n-1}^{i-j}\,,\label{ec.sn1}
\end{eqnarray} 

then, the quantization condition (\ref{ec.qz}) can be rewritten as; 

\begin{equation}\label{ec.qz1}
	d_{n} ^0c_{n-1}^0-d_{n-1}^0 c_{n} ^0=0\,.
\end{equation}

where $n$ is the overtone number.

\subsection{Computing the QNMs}

In the following Tables \ref{qnmc}, \ref{qnmg},\ref{qnmce},\ref{qnmge}, \ref{qnmcg} and \ref{qnmgg}, we present the QNMs obtained using the AIM and compare them with the sixth-order WKB results, in order to assess and demonstrate the reliability of the AIM.

To apply the AIM, a root-finding algorithm is then employed to determine the quasi-normal modes. All calculations are performed using Mathematica software, with 30 iterations used to compute the QNMs in this work.

\begin{table}[h!]
	\begin{center}
		\begin{tabular}{c c c c c c c c c }
			\hline
			\hline
			\multicolumn{9}{ c }{Scalar perturbations}\\ \hline
			\multicolumn{9}{ c}{$\omega_q=-2/3$}\\
			&&\multicolumn{3}{ c}{AIM}&\multicolumn{4}{ c}{WKB}\\\hline 
			$n$&$l$&$c=0.005$&$c=0.03$&$c=0.06$&&$c=0.005$&$c=0.03$&$c=0.06$\\
			0&1&
			0.309167-0.092276$i$	
			&0.268914-0.079457$i$	
			&0.216811-0.062764$i$	
			&
			&0.309153-0.092400$i$
			&0.268903-0.079541$i$	
			&0.216811-0.062807$i$\\
			&2&
			0.510701-0.091495$i$	
			&0.448363-0.078372$i$	
			&0.366274-0.061543$i$	
			&
			&0.510697-0.091509$i$
			&0.448361-0.078381$i$	
			&0.366273-0.061547$i$\\ 
			1&1&
			0.287150-0.286136$i$	
			&0.251283-0.245617$i$	
			&0.204247-0.193259$i$	
			&
			&0.287326-0.286501$i$
			&0.251397-0.245889$i$	
			&0.204356-0.193365$i$\\
			&2&
			0.495526-0.278262$i$	
			&0.435957-0.238112$i$	
			&0.357278-0.186670$i$	
			&
			&0.495529-0.278324$i$
			&0.435960-0.238151$i$	
			&0.357276-0.186690$i$\\
			&3&
			0.702003-0.275812$i$	
			&0.618618-0.235784$i$	
			&0.508052-0.184700$i$	
			&
			&0.702003-0.275823$i$
			&0.618618-0.235791$i$	
			&0.508053-0.184702$i$\\ 
			2&2&
			0.469479-0.474974$i$	
			&0.414442-0.405829$i$	
			&0.341393-0.317345$i$	
			&
			&0.469541-0.475169$i$
			&0.414480-0.405954$i$	
			&0.341391-0.317423$i$\\
			&3&
			0.681138-0.465959$i$	
			&0.601386-0.397986$i$	
			&0.495413-0.311257$i$	
			&
			&0.681139-0.465993$i$
			&0.601387-0.398006$i$	
			&0.495416-0.311264$i$\\
			&4&
			0.890258-0.461879$i$	
			&0.786003-0.394462$i$	
			&0.647339-0.308584$i$	
			&
			&0.890258-0.461887$i$
			&0.786002-0.394467$i$	
			&0.647341-0.308584$i$\\ 
			 \hline
			\multicolumn{9}{ c}{$\omega_q=-4/9$}\\
			&&\multicolumn{3}{ c}{AIM}&\multicolumn{4}{ c}{WKB}\\\hline 
			0&1&
			0.313129-0.093426$i$
			&0.294042-0.086857$i$	
			&0.271313-0.079034$i$	
			&
			&0.313115-0.093555$i$	
			&0.294027-0.086969$i$	
			&0.271299-0.079126$i$\\
			&2&
			0.516609-0.092727$i$
			&0.486335-0.086180$i$	
			&0.450153-0.078387$i$
			&
			&0.516605-0.092742$i$	
			&0.486332-0.086192$i$	
			&0.450150-0.078396$i$\\ 
			1&1&
			0.290540-0.289878$i$
			&0.273078-0.269432$i$	
			&0.252282-0.245065$i$
			&
			&0.290729-0.290244$i$	
			&0.273231-0.269760$i$	
			&0.252400-0.245345$i$\\
			&2&
			0.501116-0.282050$i$
			&0.471997-0.262085$i$	
			&0.437189-0.238321$i$
			&
			&0.501119-0.282115$i$	
			&0.471999-0.262138$i$	
			&0.437191-0.238361$i$\\
			&3&
			0.709775-0.279624$i$
			&0.668856-0.259828$i$	
			&0.619888-0.236272$i$
			&
			&0.709774-0.279635$i$	
			&0.668856-0.259837$i$	
			&0.619889-0.236279$i$\\ 
			2&2&
			0.474556-0.481548$i$
			&0.447322-0.447374$i$	
			&0.414757-0.406678$i$
			&
			&0.474609-0.481758$i$	
			&0.447356-0.447550$i$	
			&0.414780-0.406810$i$\\
			&3&
			0.681138-0.465959$i$	
			&0.649155-0.438928$i$	
			&0.602048-0.399030$i$	
			&
			&0.688512-0.472487$i$	
			&0.649154-0.438956$i$	
			&0.602048-0.399049$i$\\
			&4&
			0.899895-0.468336$i$
			&0.848469-0.435128$i$	
			&0.786892-0.395614$i$
			&
			&0.899895-0.468345$i$	
			&0.848469-0.435135$i$	
			&0.786891-0.395618$i$\\ 
			\hline
			\hline		\end{tabular}
		\caption{Quasinormal frequencies for the scalar perturbations for several values of the parameter $c$, with $g^2=0.20$.}
		\label{qnmc}
	\end{center}
\end{table}

\begin{table}[h!]
	\begin{center}
		\begin{tabular}{c c c c c c c c c }
			\hline
			\hline
			\multicolumn{9}{ c }{Scalar perturbations}\\ \hline
			\multicolumn{9}{ c}{$\omega_q=-2/3$}\\
			&&\multicolumn{3}{ c}{AIM}&\multicolumn{4}{ c}{WKB} 
			\\\hline 
			$n$&$l$&$g^2=0.05$&$g^2=0.20$&$g^2=0.35$&&$g^2=0.05$&$g^2=0.20$&$g^2=0.35$\\
			0&1&
			0.218010-0.068672$i$	
			&0.234764-0.068520$i$	
			&0.257542-0.065795$i$	
			&
			&0.217999-0.068720$i$
			&0.234756-0.068577$i$	
			&0.257569-0.065837$i$\\
			&2&
			0.367562-0.067314$i$	
			&0.394758-0.067311$i$	
			&0.432235-0.064865$i$	
			&
			&0.367561-0.067318$i$
			&0.394756-0.067317$i$	
			&0.432236-0.064871$i$\\ 
			1&1&
			0.201818-0.212878$i$	
			&0.220532-0.211267$i$	
			&0.244633-0.201029$i$	
			&
			&0.201852-0.213028$i$
			&0.220613-0.211452$i$	
			&0.244943-0.200961$i$\\
			&2&
			0.355946-0.204759$i$	
			&0.384608-0.204296$i$	
			&0.423485-0.196105$i$	
			&
			&0.355946-0.204773$i$
			&0.384609-0.204321$i$	
			&0.423510-0.196121$i$\\
			&3&
			0.507633-0.202292$i$	
			&0.546520-0.202177$i$	
			&0.599893-0.194617$i$	
			&
			&0.507633-0.202294$i$
			&0.546520-0.202181$i$	
			&0.599893-0.194622$i$\\ 
			2&2&
			0.335742-0.349842$i$	
			&0.366814-0.347649$i$	
			&0.407398-0.331440$i$	
			&
			&0.335704-0.349901$i$
			&0.366832-0.347734$i$	
			&0.407556-0.331420$i$\\
			&3&
			0.491405-0.341884$i$	
			&0.532313-0.340925$i$	
			&0.587525-0.326902$i$	
			&
			&0.491395-0.341887$i$
			&0.532310-0.340939$i$	
			&0.587527-0.326918$i$\\
			&4&
			0.6446770-0.338362$i$	
			&0.695642-0.337943$i$	
			&0.765245-0.324901$i$	
			&
			&0.644674-0.338362$i$
			&0.695643-0.337945$i$	
			&0.765242-0.324906$i$\\ 
		\hline
		\multicolumn{9}{ c}{$\omega_q=-4/9$}\\
		&&\multicolumn{3}{ c}{AIM}&\multicolumn{4}{ c}{WKB} 
		\\\hline 
		0&1&
		0.263387-0.083222$i$
		&0.278869-0.081634$i$	
		&0.299505-0.077017$i$
		&
		&0.263366-0.083308$i$
		&0.278854-0.081732$i$	
		&0.299503-0.077098$i$\\
		&2&
		0.436948-0.082446$i$
		&0.462198-0.080976$i$	
		&0.496638-0.076543$i$
		&
		&0.436947-0.082453$i$
		&0.462196-0.080986$i$	
		&0.496635-0.076554$i$\\ 
		1&1&
		0.240465-0.259913$i$
		&0.259196-0.253164$i$	
		&0.281758-0.236154$i$
		&
		&0.240510-0.260160$i$
		&0.259320-0.253465$i$	
		&0.281870-0.236349$i$\\
		&2&
		0.421192-0.251398$i$
		&0.448778-0.246216$i$	
		&0.485116-0.231643$i$
		&
		&0.421191-0.251423$i$
		&0.448780-0.246261$i$	
		&0.485116-0.231691$i$\\
		&3&
		0.599321-0.248798$i$
		&0.636198-0.244098$i$	
		&0.685919-0.230282$i$
		&
		&0.599321-0.248802$i$
		&0.636198-0.244106$i$	
		&0.685918-0.230291$i$\\ 
		2&2&
		0.394278-0.431208$i$
		&0.425600-0.420201$i$	
		&0.463987-0.392159$i$
		&
		&0.394199-0.431319$i$
		&0.425624-0.420350$i$	
		&0.464039-0.392295$i$\\
		&3&
		0.577707-0.421415$i$
		&0.617740-0.412285$i$	
		&0.669830-0.387147$i$
		&
		&0.577689-0.421421$i$
		&0.617740-0.412307$i$	
		&0.669833-0.387176$i$\\
		&4&
		0.758363-0.417034$i$
		&0.807406-0.408741$i$	
		&0.872751-0.384928$i$
		&
		&0.758359-0.417034$i$
		&0.807406-0.408746$i$	
		&0.872752-0.384935$i$\\ 
			\hline
			\hline		\end{tabular}
		\caption{Quasinormal frequencies for the scalar perturbations for several values of the parameter $g^2$, with $c=0.05$.}
		\label{qnmg}
	\end{center}
\end{table}

\begin{table}[h!]
	\begin{center}
		\begin{tabular}{c c c c c c c c c }
			\hline
			\hline
			\multicolumn{9}{ c }{Electromagnetic perturbations}\\ \hline
			\multicolumn{9}{ c}{$\omega_q=-2/3$}\\
			&&\multicolumn{3}{ c}{AIM}&\multicolumn{4}{ c}{WKB}\\\hline 
			$n$&$l$&$c=0.005$&$c=0.03$&$c=0.06$&&$c=0.005$&$c=0.03$&$c=0.06$\\
			0&1&
			0.267544-0.087972$i$	
			&0.237310-0.075340$i$	
			&0.196673-0.059221$i$	
			&
			&0.267362-0.088406$i$	
			&0.237203-0.075608$i$	
			&0.196647- 0.05934$i$\\
			&2&
			0.486274-0.089965$i$	
			&0.429941-0.076911$i$	
			&0.354635-0.060301$i$	
			&
			&0.486268-0.089985$i$	
			&0.429938-0.076923$i$	
			&0.354632-0.060307$i$\\ 
			1&1&
			0.241782-0.275358$i$	
			&0.215815-0.235231$i$	
			&0.180742-0.183962$i$	
			&
			&0.241930-0.277143 $i$	
			&0.215898-0.236342$i$	
			&0.180935-0.184323$i$\\
			&2&
			0.470246-0.273949$i$	
			&0.416664-0.233971$i$	
			&0.344932-0.183080$i$	
			&
			&0.470246-0.274042$i$	
			&0.416663-0.234027$i$	
			&0.344922-0.183111$i$\\
			&3&
			0.684337-0.273542$i$	
			&0.605237-0.233612$i$	
			&0.499566-0.182842$i$	
			&
			&0.684336-0.273555$i$	
			&0.605236-0.233620$i$	
			&0.499564-0.182846$i$\\ 
			2&2&
			0.442854-0.468627$i$	
			&0.393686-0.399709$i$	
			&0.327725-0.311890$i$
			&
			&0.442888-0.468916$i$	
			&0.393691-0.399890$i$	
			&0.327663-0.312015$i$\\
			&3&
			0.662908-0.462409$i$	
			&0.587411-0.394575$i$	
			&0.486429-0.308284$i$
			&
			&0.662906-0.462448$i$	
			&0.587406-0.394599$i$	
			&0.486418-0.308298$i$\\
			&4&
			0.876364-0.459646$i$	
			&0.775435-0.392324$i$	
			&0.640608-0.306743$i$
			&
			&0.876363-0.459655$i$	
			&0.775435-0.392329$i$	
			&0.640610-0.306743$i$\\ 
			\hline
			\multicolumn{9}{ c}{$\omega_q=-4/9$}\\
			&&\multicolumn{3}{ c}{AIM}&\multicolumn{4}{ c}{WKB}\\\hline 
			0&1&
			0.270305-0.089183$i$	
			&0.255345-0.082973$i$	
			&0.237393-0.0755695$i$
			&
			&0.270118-0.089636$i$	
			&0.255197-0.083336$i$	
			&0.237285-0.075842$i$\\
			&2&
			0.491454-0.091221$i$	
			&0.463637-0.084806$i$	
			&0.430291-0.0771676$i$
			&
			&0.491447-0.091241$i$	
			&0.463632-0.084822$i$	
			&0.430287-0.077179$i$\\ 
			1&1&
			0.244099-0.279225$i$	
			&0.231036-0.259619$i$	
			&0.215356-0.236235$i$
			&
			&0.244285-0.281054$i$	
			&0.231172-0.261103$i$	
			&0.215453-0.237355$i$\\
			&2&
			0.475140-0.277801$i$	
			&0.448546-0.258200$i$	
			&0.416657-0.234858$i$
			&
			&0.475140-0.277898$i$	
			&0.448546-0.258276$i$	
			&0.416656-0.234915$i$\\
			&3&
			0.691597-0.277389$i$	
			&0.652457-0.257790$i$	
			&0.605542-0.234461$i$
			&
			&0.691597-0.277402$i$	
			&0.652457-0.257800$i$	
			&0.605542-0.234469$i$\\ 
			2&2&
			0.447289-0.475286$i$	
			&0.422658-0.441628$i$	
			&0.393118-0.401528$i$
			&
			&0.447328-0.475586$i$	
			&0.422680-0.441867$i$	
			&0.393118-0.401711$i$\\
			&3&
			0.669793-0.468954$i$	
			&0.632256-0.435730$i$	
			&0.587252-0.396179$i$
			&
			&0.669793-0.468994$i$	
			&0.632254-0.435762$i$	
			&0.587249-0.396202$i$\\
			&4&
			0.885611-0.466138$i$	
			&0.835581-0.433123$i$	
			&0.775616-0.393830$i$
			&
			&0.885610-0.466147$i$	
			&0.835580-0.433130$i$	
			&0.775614-0.393836$i$\\ 
			\hline
			\hline		\end{tabular}
		\caption{Quasinormal frequencies for the electromagnetic perturbations for several values of the parameter $c$, with $g^2=0.20$.}
		\label{qnmce}
	\end{center}
\end{table}

\begin{table}[h!]
	\begin{center}
		\begin{tabular}{c c c c c c c c c }
			\hline
			\hline
			\multicolumn{9}{ c }{Electromagnetic perturbations}\\ \hline
			\multicolumn{9}{ c}{$\omega_q=-2/3$}\\
			&&\multicolumn{3}{ c}{AIM}&\multicolumn{4}{ c}{WKB} 
			\\\hline 
			$n$&$l$&$g^2=0.05$&$g^2=0.20$&$g^2=0.35$&&$g^2=0.05$&$g^2=0.20$&$g^2=0.35$\\
			0&1&
			0.194661-0.064496$i$	
			&0.210888-0.064736$i$	
			&0.233733-0.062439$i$	
			&
			&0.194630-0.064584$i$
			&0.210814-0.064906$i$	
			&0.233636-0.062646$i$\\
			&2&
			0.354108-0.065869$i$	
			&0.380919-0.065977$i$	
			&0.418281-0.063654$i$	
			&
			&0.354107-0.065873$i$
			&0.380917-0.065985$i$	
			&0.418276-0.063664$i$\\ 
			1&1&
			0.174129-0.202198$i$	
			&0.193036-0.201506$i$	
			&0.218718-0.191935$i$	
			&
			&0.174110-0.202486$i$
			&0.193007-0.202280$i$	
			&0.218674-0.192861$i$\\
			&2&
			0.341546-0.200619$i$	
			&0.369985-0.200472$i$	
			&0.409076-0.192588$i$	
			&
			&0.341543-0.200634$i$
			&0.369981-0.200509$i$	
			&0.409062-0.192638$i$\\
			&3&
			0.497783-0.200147$i$	
			&0.536440-0.200187$i$	
			&0.589819-0.192789$i$	
			&
			&0.497783-0.200149$i$
			&0.536439-0.200191$i$	
			&0.589819-0.192796$i$\\ 
			2&2&
			0.319602-0.343686$i$	
			&0.350781-0.341911$i$	
			&0.392203-0.325959$i$	
			&
			&0.319500- 0.34375$i$
			&0.350751-0.342041$i$	
			&0.392177-0.326133$i$\\
			&3&
			0.480887-0.338484$i$	
			&0.521686-0.337766$i$	
			&0.577143-0.323955$i$	
			&
			&0.480870-0.336848$i$
			&0.521679-0.337781$i$	
			&0.577148-0.323972$i$\\
			&4&
			0.636837-0.336247$i$	
			&0.687656-0.335976$i$	
			&0.757329-0.323080$i$	
			&
			&0.636833-0.336246$i$
			&0.687655-0.335979$i$	
			&0.757328-0.323085$i$\\ 
			\hline
			\multicolumn{9}{ c}{$\omega_q=-4/9$}\\
			&&\multicolumn{3}{ c}{AIM}&\multicolumn{4}{ c}{WKB} 
			\\\hline 
			0&1&
			0.227443-0.079155$i$	
			&0.243379-0.078031$i$	
			&0.265469-0.073697$i$	
			&
			&0.227381-0.079314$i$
			&0.243259-0.078331$i$	
			&0.265298-0.074056$i$\\
			&2&
			0.415997-0.081053$i$	
			&0.441405-0.079706$i$	
			&0.476504-0.075334$i$	
			&
			&0.415995-0.081060$i$
			&0.441401-0.079719$i$	
			&0.476498-0.075350$i$\\ 
			1&1&
			0.200516-0.249761$i$	
			&0.220585-0.244009$i$	
			&0.246321-0.227040$i$	
			&
			&0.200458-0.250340$i$
			&0.220694-0.245243$i$	
			&0.246307-0.228586$i$\\
			&2&
			0.399317-0.247472$i$	
			&0.427286-0.242615$i$	
			&0.464677-0.228126$i$	
			&
			&0.399314-0.247498$i$
			&0.427286-0.242678$i$	
			&0.464678-0.228202$i$\\
			&3&
			0.584128-0.246753$i$	
			&0.621178-0.242212$i$	
			&0.671474-0.228449$i$	
			&
			&0.584128-0.246756$i$
			&0.621178-0.242221$i$	
			&0.671474-0.228460$i$\\ 
			2&2&
			0.370853-0.425494$i$	
			&0.402965-0.414855$i$	
			&0.443097-0.386618$i$	
			&
			&0.370710- 0.425611$i$
			&0.402970-0.415058$i$	
			&0.443179-0.386816$i$\\
			&3&
			0.561881-0.418225$i$	
			&0.602252-0.409320$i$	
			&0.655195-0.384184$i$	
			&
			&0.561858-0.418230$i$
			&0.602249-0.409346$i$	
			&0.655206-0.384214$i$\\
			&4&
			0.746377-0.415033$i$	
			&0.795601-0.406885$i$	
			&0.861469-0.383098$i$	
			&
			&0.746372-0.415033$i$
			&0.795600-0.406890$i$	
			&0.861470-0.383106$i$\\ 
			\hline
			\hline		\end{tabular}
		\caption{Quasinormal frequencies for the electromagnetic perturbations for several values of the parameter $g^2$, with $c=0.05$.}
		\label{qnmge}
	\end{center}
\end{table}

\begin{table}[h!]
	\begin{center}
		\begin{tabular}{c c c c c c c c c }
			\hline
			\hline
			\multicolumn{9}{ c }{Gravitational perturbations}\\ \hline
			\multicolumn{9}{ c}{$\omega_q=-2/3$}\\
			&&\multicolumn{3}{ c}{AIM}&\multicolumn{4}{ c}{WKB}\\\hline 
			$n$&$l$&$c=0.005$&$c=0.03$&$c=0.06$&&$c=0.005$&$c=0.03$&$c=0.06$\\
			0&2&
			0.396864-0.084396$i$	
			&0.350395-0.072833$i$	
			&0.288579-0.057824$i$	
			&
			&0.396866-0.084450$i$	
			&0.350394-0.072864$i$	
			&0.288579-0.057837$i$\\
			&3&
			0.634968-0.087934$i$	
			&0.560853-0.075463$i$	
			&0.461982-0.059464$i$	
			&
			&0.634967-0.087939$i$	
			&0.560853-0.075465$i$	
			&0.461982-0.059466$i$\\ 
			1&2&
			0.376881-0.258387$i$	
			&0.334215-0.222604$i$	
			&0.277053-0.176231$i$	
			&
			&0.376955-0.258778$i$	
			&0.334244-0.222822$i$	
			&0.277075-0.176306$i$\\
			&3&
			0.622198-0.266121$i$	
			&0.550408-0.228222$i$	
			&0.454469-0.179618$i$	
			&
			&0.622195-0.266146$i$	
			&0.550407-0.228236$i$	
			&0.454465-0.179625$i$\\
			&4&
			0.846596-0.269072$i$	
			&0.748246-0.230341$i$	
			&0.616949-0.180878$i$	
			&
			&0.846595-0.269076$i$	
			&0.748246-0.230344$i$	
			&0.616949-0.180879$i$\\ 
			2&2&
			0.343051-0.446310$i$	
			&0.306533-0.383441$i$	
			&0.256969-0.302224$i$	
			&
			&0.343284-0.447637$i$	
			&0.306597-0.384201$i$	
			&0.256990-0.302455$i$\\
			&3&
			0.598586-0.450900$i$	
			&0.530981-0.386225$i$	
			&0.440331-0.303320$i$	
			&
			&0.598574-0.450970$i$	
			&0.530976-0.386261$i$	
			&0.440309-0.303346$i$\\
			&4&
			0.828286-0.452767$i$	
			&0.733170-0.387333$i$	
			&0.605996-0.303770$i$	
			&
			&0.828284-0.452778$i$	
			&0.733170-0.387338$i$	
			&0.605996-0.303772$i$\\ 
			\hline
			\multicolumn{9}{ c}{$\omega_q=-4/9$}\\
			&&\multicolumn{3}{ c}{AIM}&\multicolumn{4}{ c}{WKB}\\\hline 
			0&2&
			0.401167-0.085463$i$	
			&0.378276-0.079653$i$	
			&0.350870-0.072709$i$	
			&
			&0.401171-0.085519$i$	
			&0.378280-0.079694$i$	
			&0.350873-0.072736$i$\\
			&3&
			0.641800-0.089115$i$	
			&0.605230-0.082934$i$	
			&0.561413-0.0755632$i$	
			&
			&0.641799-0.089119$i$	
			&0.605229-0.082938$i$	
			&0.561412-0.075565$i$\\ 
			1&2&
			0.380744-0.261716$i$	
			&0.359432-0.243841$i$	
			&0.333908-0.222472$i$	
			&
			&0.380825-0.262123$i$	
			&0.359495-0.244133$i$	
			&0.333947-0.222667$i$\\
			&3&
			0.628775-0.269716$i$	
			&0.593207-0.250970$i$	
			&0.550581-0.228610$i$	
			&
			&0.628773-0.269742$i$	
			&0.593206-0.250989$i$	
			&0.550579-0.228625$i$\\
			&4&
			0.855635-0.272771$i$	
			&0.807006-0.253669$i$	
			&0.748731-0.230910$i$	
			&
			&0.855635-0.272775$i$	
			&0.807006-0.253672$i$	
			&0.748731-0.230913$i$\\ 
			2&2&
			0.346229-0.452256$i$	
			&0.327466-0.421168$i$	
			&0.304986-0.383993$i$	
			&
			&0.346452-0.453632$i$	
			&0.327583-0.422166$i$	
			&0.304994-0.384680$i$\\
			&3&
			0.604706-0.457059$i$	
			&0.570940-0.425179$i$	
			&0.530460-0.387153$i$	
			&
			&0.604693-0.457134$i$	
			&0.570928-0.425235$i$	
			&0.530447-0.387192$i$\\
			&4&
			0.836979-0.459027$i$	
			&0.789762-0.426813$i$	
			&0.733168-0.388430$i$	
			&
			&0.836976-0.459039$i$	
			&0.789760-0.426822$i$	
			&0.733165-0.388437$i$\\ 
			\hline
			\hline		\end{tabular}
		\caption{Quasinormal frequencies for the gravitational perturbations for several values of the parameter $c$, with $g^2=0.20$.}
		\label{qnmcg}
	\end{center}
\end{table}

\begin{table}[h!]
	\begin{center}
		\begin{tabular}{c c c c c c c c c }
			\hline
			\hline
			\multicolumn{9}{ c }{Gravitational perturbations}\\ \hline
			\multicolumn{9}{ c}{$\omega_q=-2/3$}\\
			&&\multicolumn{3}{ c}{AIM}&\multicolumn{4}{ c}{WKB} 
			\\\hline 
			$n$&$l$&$g^2=0.05$&$g^2=0.20$&$g^2=0.35$&&$g^2=0.05$&$g^2=0.20$&$g^2=0.35$\\
			0&2&
			0.288166-0.062929$i$	
			&0.310118-0.062993$i$	
			&0.340655-0.060741$i$	
			&
			&0.288174-0.062909$i$	
			&0.310115-0.063011$i$	
			&0.340620-0.060799$i$\\
			&3&
			0.462220-0.064840$i$	
			&0.496464-0.064949$i$	
			&0.544067-0.062727$i$	
			&
			&0.462220-0.064841$i$	
			&0.496463-0.064950$i$	
			&0.544066-0.062730$i$\\ 
			1&2&
			0.273003-0.192520$i$	
			&0.297024-0.192190$i$	
			&0.329894-0.184327$i$	
			&
			&0.273008-0.192372$i$	
			&0.297019-0.192322$i$	
			&0.329801-0.184699$i$\\
			&3&
			0.452463-0.196199$i$	
			&0.487951-0.196278$i$	
			&0.536928-0.189113$i$	
			&
			&0.452464-0.196199$i$	
			&0.487950-0.196286$i$	
			&0.536927-0.189124$i$\\
			&4&
			0.616084-0.197637$i$	
			&0.662752-0.197805$i$	
			&0.727304-0.190787$i$
			&
			&0.616083-0.197638$i$	
			&0.662753-0.197807$i$	
			&0.727304-0.190790$i$\\ 
			2&2&
			0.246782-0.332460$i$	
			&0.274379-0.330144$i$	
			&0.310497-0.313594$i$	
			&
			&0.246403-0.332038$i$	
			&0.274300-0.330646$i$	
			&0.310540-0.314769$i$\\
			&3&
			0.434199-0.332434$i$	
			&0.472009-0.331730$i$	
			&0.523297-0.318162$i$	
			&
			&0.434181-0.332423$i$	
			&0.471999-0.331754$i$	
			&0.523301-0.318194$i$\\
			&4&
			0.601937-0.332571$i$	
			&0.650388-0.332365$i$	
			&0.716782-0.319715$i$	
			&
			&0.601928-0.332572$i$	
			&0.650394-0.332365$i$	
			&0.716779-0.319722$i$\\ 
			\hline
			\multicolumn{9}{ c}{$\omega_q=-4/9$}\\
			&&\multicolumn{3}{ c}{AIM}&\multicolumn{4}{ c}{WKB} 
			\\\hline 
			0&2&
			0.339286-0.076272$i$	
			&0.360000-0.075020$i$	
			&0.388529-0.070949$i$	
			&
			&0.339294-0.076228$i$	
			&0.360004-0.075051$i$	
			&0.388422-0.071084$i$\\
			&3&
			0.543881-0.079281$i$	
			&0.576014-0.078014$i$	
			&0.620375-0.0738851$i$	
			&
			&0.543881-0.079282$i$	
			&0.576013-0.078017$i$	
			&0.620374-0.073889$i$\\ 
			1&2&
			0.318240-0.234244$i$	
			&0.342413-0.229583$i$	
			&0.374361-0.215676$i$	
			&
			&0.318192-0.233950$i$	
			&0.342464-0.229804$i$	
			&0.373923-0.216688$i$
			\\
			&3&
			0.530632-0.240278$i$	
			&0.564787-0.236045$i$	
			&0.611071-0.222895$i$	
			&
			&0.530632-0.240279$i$	
			&0.564785-0.236061$i$	
			&0.611071-0.222914$i$\\
			&4&
			0.723861-0.242655$i$	
			&0.768152-0.238475$i$	
			&0.828641-0.225433$i$	
			&
			&0.723861-0.242655$i$	
			&0.768151-0.238477$i$	
			&0.828640-0.225437$i$\\ 
			2&2&
			0.282401-0.407171$i$	
			&0.312479-0.396365$i$	
			&0.348961-0.367979$i$	
			&
			&0.281438-0.406454$i$	
			&0.312535-0.397122$i$	
			&0.348266-0.371760$i$\\
			&3&
			0.506085-0.408292$i$	
			&0.543952-0.399797$i$	
			&0.593360-0.375432$i$	
			&
			&0.506040-0.408281$i$	
			&0.543938-0.399842$i$	
			&0.593376-0.375481$i$\\
			&4&
			0.704900-0.408999$i$	
			&0.752030-0.401187$i$	
			&0.815001-0.378016$i$	
			&
			&0.704893-0.408996$i$	
			&0.752028-0.401190$i$	
			&0.815000-0.378027$i$\\ 
			\hline
			\hline		\end{tabular}
		\caption{Quasinormal frequencies for the gravitational perturbations for several values of the parameter $g^2$, with $c=0.05$.}
		\label{qnmgg}
	\end{center}
\end{table}

\begin{figure}[h!]
	\centering
	\includegraphics[scale=0.9]{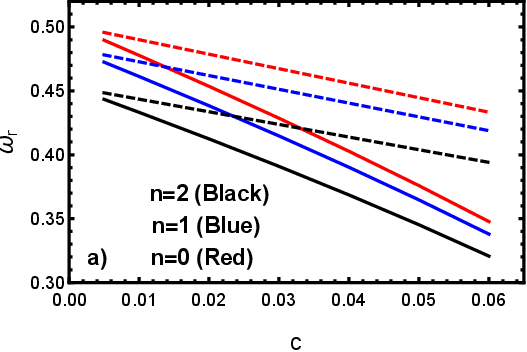}
	\includegraphics[scale=0.89]{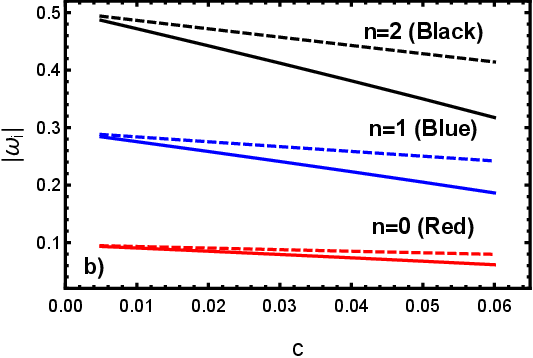}
	\caption{a) The behavior of $\omega_r$ as a function of $c$ is shown for $\omega_q = -2/3$ and $\omega_q = -4/9$ for the scalar perturbation.
		b) The behavior of $|\omega_i|$ as a function of $c$ is shown for the same values of $\omega_q$. In both cases, the remaining parameters are fixed at $g^2 = 0.1$ and $l = 2$. The solid and dashed lines correspond to $\omega_q = -2/3$ and $\omega_q = -4/9$, respectively.}
	\label{fwc}
\end{figure}

\begin{figure}[h!]
	\centering
	\includegraphics[scale=0.9]{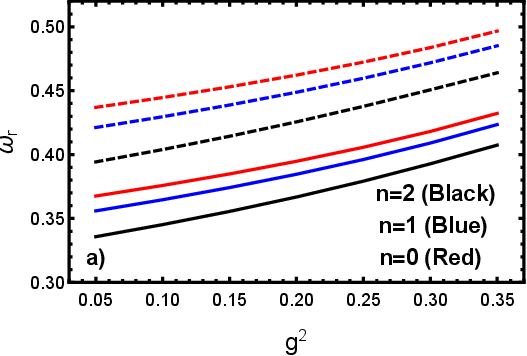}
	\includegraphics[scale=0.89]{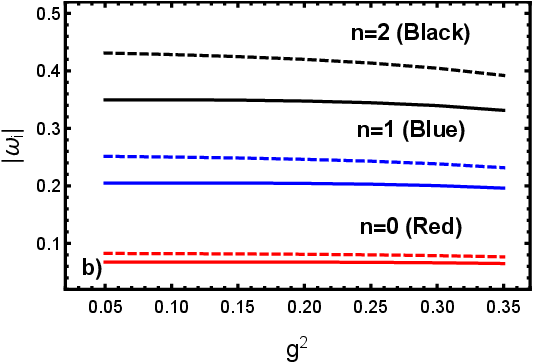}
	\caption{a) The behavior of $\omega_r$ as a function of $g^2$ is shown for $\omega_q = -2/3$ and $\omega_q = -4/9$ for the scalar perturbation.
		b) The behavior of $|\omega_i|$ as a function of $g^2$ is shown for the same values of $\omega_q$. In both cases, the remaining parameters are fixed at $c = 0.05$ and $l = 2$. The solid and dashed lines correspond to $\omega_q = -2/3$ and $\omega_q = -4/9$, respectively.}
	\label{fwg}
\end{figure}

In Tables \ref{qnmc} and \ref{qnmg}, we present the QNMs for scalar perturbations corresponding to $\omega_q = -2/3$ and $\omega_q = -4/9$, respectively, for different values of the parameters $c$ and $g^{2}$. The behavior of the QNMs for $l=2$ as functions of the parameters $c$ and $g^2$ is shown in Figs.~ \ref{fwc}–\ref{fwg}. From the Fig ~\ref{fwc}  with $g^2$ fixe, it can be observed that both $\omega_r$ and $|\omega_i|$ are significantly larger for $\omega_q=-4/9$ than for $\omega_q=-2/3$. The same behavior is observed when $c$ is kept fixed and $g^{2}$ is allowed to vary (see Fig ~\ref{fwg}).


\begin{figure}[h!]
	\centering
	\includegraphics[scale=0.9]{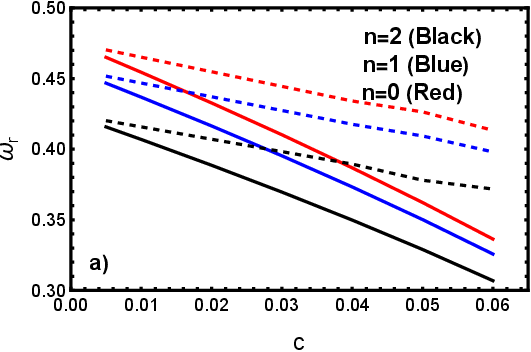}
	\includegraphics[scale=0.89]{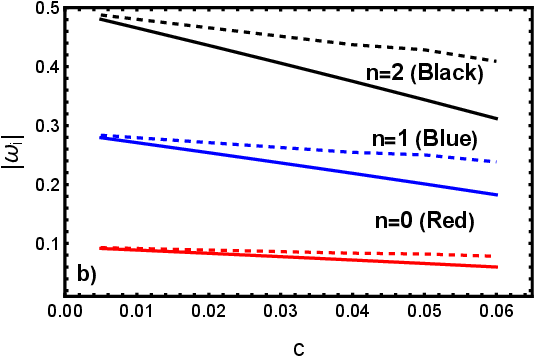}
	\caption{a) Real part of the quasinormal frequency, $\omega_r$, as a function of $c$ for the electromagnetic perturbation with $\omega_q = -2/3$ and $\omega_q = -4/9$.
		b) Absolute value of the imaginary part, $|\omega_i|$, as a function of $c$ for the same values of $\omega_q$. In both panels, the parameters are fixed at $g^2 = 0.1$ and $l = 2$. The solid and dashed lines correspond to $\omega_q = -2/3$ and $\omega_q = -4/9$, respectively.}
	\label{fwce}
\end{figure}

The QNMs for electromagnetic perturbations are presented in Tables \ref{qnmce} and \ref{qnmge} for the cases $\omega_q = -2/3$ and $\omega_q = -4/9$, respectively. From Fig.~\ref{fwce}, we observe the behavior of $\omega_{r}$ and $|\omega_{i}|$ for fixed $g^{2}$ and varying $c$. In both cases, the frequencies decrease as $c$ increases. 

\begin{figure}[h!]
	\centering
	\includegraphics[scale=0.9]{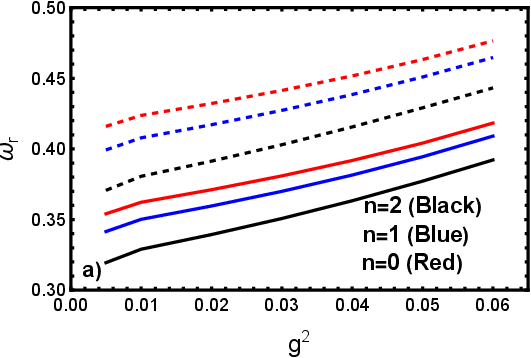}
	\includegraphics[scale=0.89]{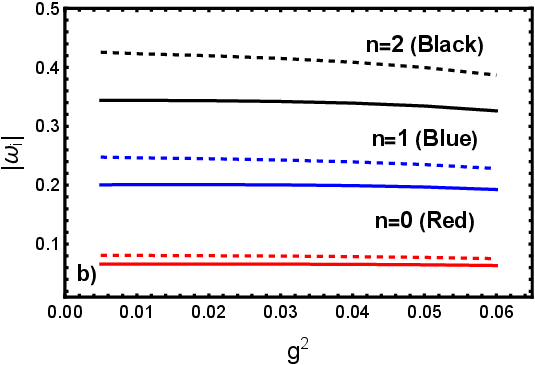}
	\caption{a) Real part of the quasinormal frequency, $\omega_r$, as a function of $g^2$ for the electromagnetic perturbation with $\omega_q = -2/3$ and $\omega_q = -4/9$.
		b) Absolute value of the imaginary part, $|\omega_i|$, as a function of $g^2$ for the same values of $\omega_q$. In both panels, the parameters are fixed at $c = 0.05$ and $l = 2$. The solid and dashed lines correspond to $\omega_q = -2/3$ and $\omega_q = -4/9$, respectively.}
	\label{fwge}
\end{figure}

This behavior indicates that the presence of the quintessence field, weakens the effective potential. But when $c$ is fixed and $g^{2}$ varying the oscillation frequency increased (see Fig \ref{fwge} a)).  On the other hand, as $|\omega_{i}|$ decreases, the magnitude of the damping rate is also reduced. The same qualitative behavior is observed when $c$ is kept fixed and $g^{2}$ is varied (see Fig.~\ref{fwge} b)).


\begin{figure}[h!]
	\centering
	\includegraphics[scale=0.9]{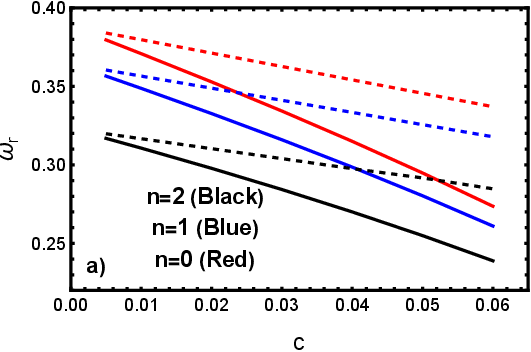}
	\includegraphics[scale=0.89]{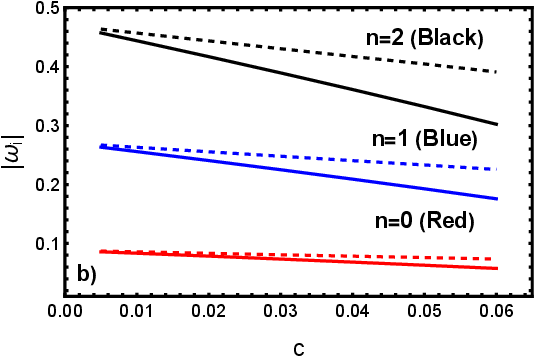}
	\caption{a) Real part of the quasinormal frequency, $\omega_r$, as a function of $c$ for gravitational perturbation with $\omega_q = -2/3$ and $\omega_q = -4/9$.
		b) Absolute value of the imaginary part, $|\omega_i|$, as a function of $c$ for the same values of $\omega_q$. In both panels, the parameters are fixed at $g^2 = 0.1$ and $l = 2$. The solid and dashed lines correspond to $\omega_q = -2/3$ and $\omega_q = -4/9$, respectively.}
	\label{fwcg}
\end{figure}

\begin{figure}[h!]
	\centering
	\includegraphics[scale=0.9]{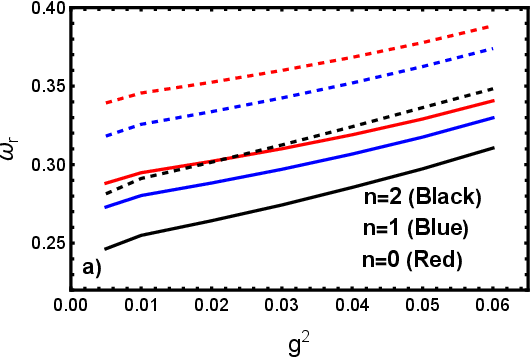}
	\includegraphics[scale=0.89]{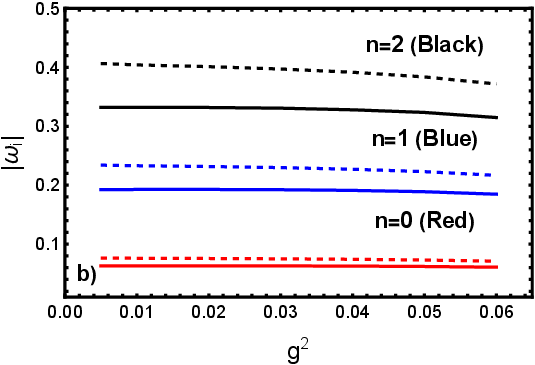}
	\caption{a) Real part of the quasinormal frequency, $\omega_r$, as a function of $g^2$ for gravitational perturbation with $\omega_q = -2/3$ and $\omega_q = -4/9$.
		b) Absolute value of the imaginary part, $|\omega_i|$, as a function of $g^2$ for the same values of $\omega_q$. In both panels, the parameters are fixed at $c = 0.05$ and $l = 2$. The solid and dashed lines correspond to $\omega_q = -2/3$ and $\omega_q = -4/9$, respectively.}
	\label{fwgg}
\end{figure}

Finally, the QNMs for gravitational perturbations are presented in Tables \ref{qnmcg} and \ref{qnmgg}, using the same parameter values as in the scalar and electromagnetic cases. It can be observed that the overall behavior of the QNMs is very similar to that found for the other types of perturbations. In particular, both $\omega_{r}$ and $|\omega_{i}|$ decrease as $c$ increases for $\omega_q = -2/3$ and $\omega_q = -4/9$ (see Fig.~\ref{fwcg}). However, the decrease in the QNMs is noticeably slower for the case $\omega_q = -4/9$. The behavior of QNMs with fixed $c$ and $g^{2}$ in the case of gravitational permutations is also observed in Fig ~\ref{fwgg}.

In general, we observe that the values of $\omega_{r}$ decrease as the quintessence parameter increases. In other words, the presence of quintessence suppresses the oscillatory behavior of the perturbations, regardless of the overtone number $n$.  This behavior can be understood from the modification of the effective potential induced by the quintessence field. 

One of the important properties is the relaxation time, defined as the inverse of the imaginary part of the QNMs ($1/|\omega_{i}|$). Therefore, in the presence of quintessence, the relaxation time (or damping time) increases for all perturbations, indicating that the black hole returns to equilibrium more slowly when surrounded by quintessence.


\clearpage
\section{Conclusions}\label{conclu}

In this work, we analysed the QNMs of scalar, electromagnetic and gravitational perturbations of  the Ay\'on–Beato–Garc\'ia (ABG) regular black hole immersed in a quintessence background. The analysis of the horizon structure showed that the quintessence parameter $c$ and the magnetic charge $g^{2}$ significantly modify the number and location of the horizons, introducing an additional cosmological horizon in a wide region of the parameter space.
 
The effective potential associated with scalar, electromagnetic and gravitational perturbations exhibited a strong dependence on both the  parameter $c$ and the quintessence state parameter $\omega_q$. In particular, increasing $c$ lowers the potential barrier, while the value of $\omega_q $ determines the relative magnitude of this reduction $V_{\omega_q=-4/9} > V_{\omega_q=-2/3}$. This behaviour directly impacts the dynamics of perturbations.

Using the improved Asymptotic Iteration Method with 30 iterations, we obtained numerically stable and accurate QNMs, in agreement with 
sixth-order WKB calculations. This agreement demonstrates the robustness and reliability of AIM for analyzing the QNMs of black holes. Moreover, 
for this particular black hole, the computation time required to obtain the QNMs was shorter using AIM than with the WKB approach.

In the  Tables \ref{qnmc}, \ref{qnmg},\ref{qnmce},\ref{qnmge}, \ref{qnmcg} and \ref{qnmgg} exhibit a clear dependence of the QNMs on $c$ and $\omega_{q}$. Increasing $c$ reduces both $\omega_{r}$ and $|\omega_{i}|$, leading to slower damping of the perturbations, whereas larger $g^{2}$ increases the oscillation frequency and reduces the damping rate. These trends are consistent with previous studies of black holes in dark-energy backgrounds.

Overall, our results indicate that quintessence softens the dynamical response of the ABG black hole and increases the relaxation time of perturbations. We emphasize  the importance of considering more realistic black hole models in which quintessence plays a significant role in their dynamics. Such scenarios are particularly relevant since the results may have observational implications for the analysis of gravitational wave signals, offering a theoretical framework that links oscillation spectra to the presence of dark energy. In this context, the modification of the metric directly influences the behaviour of the oscillations.

\section*{ACKNOWLEDGMENT}

The authors acknowledge partial support from SNII–SECIHTI, Mexico. Diego Ariel Sotelo also acknowledges the Mesoamerican Center for Theoretical Physics.

\bibliographystyle{unsrt}
\bibliography{bibliografia}

\end{document}